\documentclass[9pt]{article}

\usepackage[pdftex]{graphicx,color}

\begin{document} 

\def\gsim {>\kern-1.2em\lower1.1ex\hbox{$\sim$}~}   
\def\lsim {<\kern-1.2em\lower1.1ex\hbox{$\sim$}~}   

\pagenumbering{arabic}
\noindent
\line(1,0){345}\\*[0.08cm]
{\bf \large \noindent Cosmic Ray Acceleration in Active Galactic Nuclei}\\*[0.08cm] 
{\bf \large - On Centaurus~A as a possible UHECR Source.}\\
\line(1,0){345}\\*[0.08cm]
{\bf Frank M. Rieger}\\
{\small \it %
(1) Max-Planck-Institut f\"ur Kernphysik, 69117 Heidelberg, Germany\\
(2) European Associated Laboratory for Gamma-Ray Astronomy (CNRS/MPG)
}

     \baselineskip=10pt
     \parindent=10pt

\section*{\it \large Abstract} 
We discuss a representative selection of particle acceleration mechanisms believed 
to be operating in Active Galactic Nuclei. Starting from direct electrostatic field acceleration 
in the vicinity of the black hole up to Fermi-type particle acceleration in the jet and beyond, 
possible efficiency constraints on the energization of ultra-high energy cosmic rays (UHECR) 
are evaluated. When paradigmatically applied to Cen~A, the following results are obtained:
(i) Proton acceleration to energies of $E_c = 5\times 10^{19}$ eV and beyond remains 
challenging and most likely requires the operation of an additional mechanism capable of 
boosting energetic seed protons up by a factor of $\sim$ten. It is argued that shear 
acceleration along the large-scale jet in Cen~A could be a promising candidate for this. 
(ii) Heavier elements, like iron nuclei, are more easily accelerated (by, e.g., shocks or direct 
electrostatic fields) and may not need additional boosting to reach $E \gsim E_c$; (iii) If 
Cen~A indeed proves to be an UHECR source, the cosmic ray composition might thus be 
expected to become heavier above energies of a few times $10^{19}$ eV. \\

\section{\large Introduction}
The observation of variable, non-thermal high emission from Active Galactic Nuclei 
(AGN) reveals that efficient particle acceleration can take place on different length 
scales. It is widely believed, for example, that diffusive shock acceleration of electrons 
can produce the power-law particle distributions that are needed to account for the 
observed nuclear synchrotron and inverse Compton emission features in AGN jets. 
While efficient electron acceleration is in most cases strongly limited by radiative 
losses, this is much less the case for protons and heavier nuclei, suggesting that these 
particles could reach much higher energy via the same acceleration process. Motivated 
by the indication of a possible correlation between the Pierre Auger (PAO)-measured 
ultra-high energy cosmic ray (UHECR) events and the nearby AGN distribution 
\cite{abr07,rou09,hag09}, this contribution analyzes the conditions under which 
efficient cosmic ray acceleration to UHECR energies may become possible. Particular 
attention is given to the radio galaxy Cen~A, which, based on its proximity, could 
represent a promising UHECR source candidate, e.g., \cite{rom96,gur08,kac09}.\\
 
 \section{\large Centaurus A}
Given the possible association of some of the PAO measured UHECR events with 
Centaurus~A (Cen~A) \cite{hag09}, an application to it may appear most instructive. 
Being the nearest ($d\sim 3.4$ Mpc) FR~I source, Cen~A is among the best studied AGN. 
Radio observations show a complex morphology with a sub-pc-scale jet and counter-jet, 
a one-sided kpc-jet, two radio lobes and extended diffusive emission. VLBI observations 
suggest that Cen~A is a non-blazar source with its jet inclined at a rather large viewing 
angle $i \gsim 50^{\circ}$ and characterized by a relatively modest bulk flow speed $u_j 
\sim 0.5$ c \cite{tin98,har03}.
The center of its activity is a supermassive black hole with mass inferred to be in the range 
$m_{\rm BH} = (0.5-3) \times 10^8 M_{\odot}$ \cite{mar06,neu07}. With a bolometric luminosity 
output of the order of $L_b \sim 10^{43}$ erg/s \cite{why04}, Cen~A is rather under-luminous 
and accreting at sub-Eddington rates. If the inner disk in Cen~A remains cooling-dominated 
(standard disk), accreting rates $\dot{m}\sim 10^{-3} \dot{m}_{\rm Edd}$ and equipartition 
magnetic field strengths close to the black hole of order $B_0\sim (2 L_b/r_g^2 c)^{1/2} \sim 
2 \times 10^3$ G might be expected (where $r_g = GM/c^2 \simeq 1.5 \times 10^{13}$ 
cm is the gravitational radius for a $10^8 M_{\odot}$ black hole). If the disk switches to a 
radiatively inefficient mode, characteristic magnetic field strengths may be somewhat 
higher, possibly reaching $B_0 \sim 2\times 10^4$ G.\\

\begin{figure}[t]
  \begin{center}
   \includegraphics[angle=270,width=13.3cm]{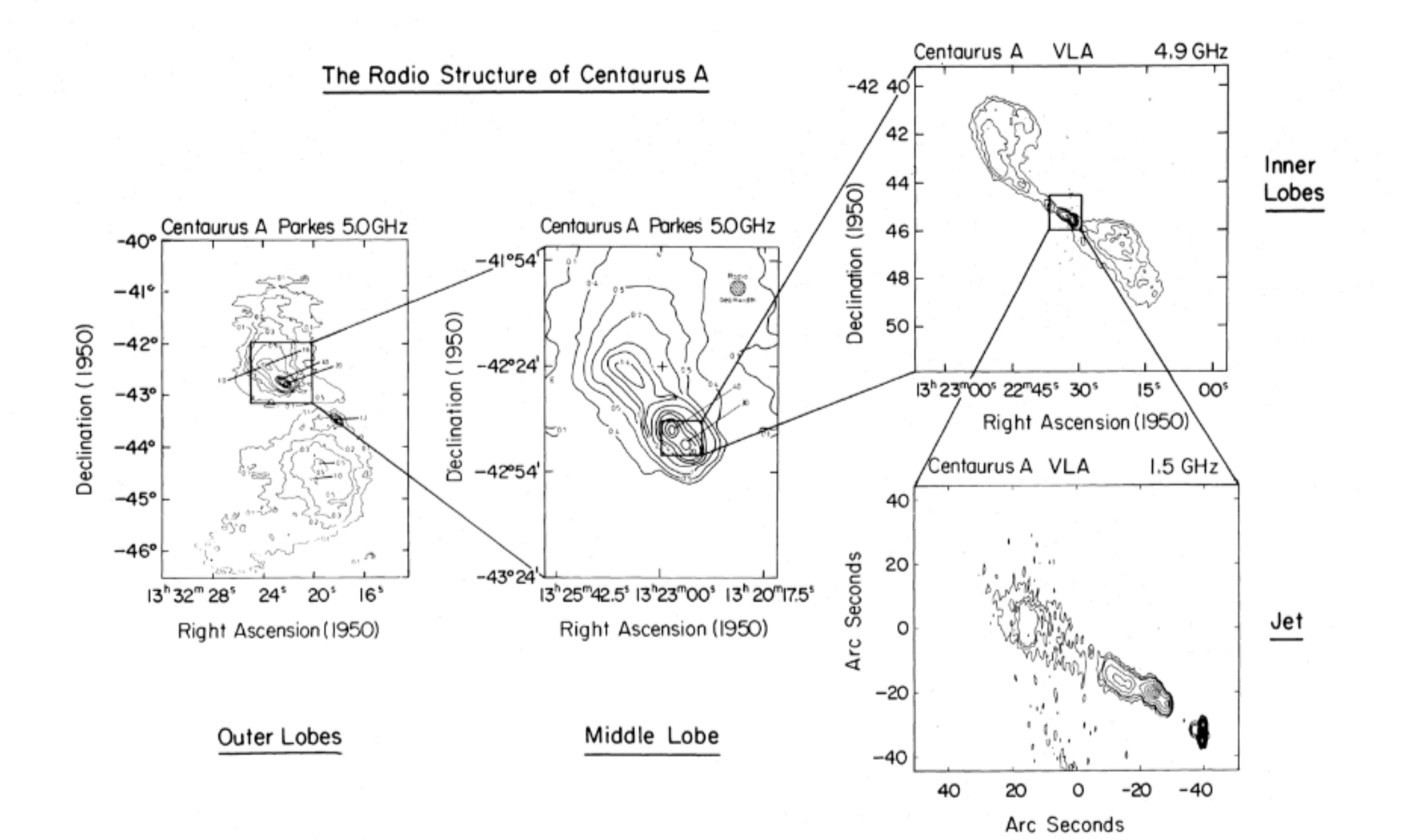}
   \end{center}
  \caption{The radio morphology of Cen~A, including the outer lobes (scale: several 
  hundred kpc), the middle lobes (scale: several tens of kpc) and the large-scale jet 
  (scale: several kpc). From Burns et al.~\cite{bur83}. (Reproduced by permission of 
  the AAS).}
\end{figure}

\section{\large Particle acceleration in the vicinity of the black hole}
Rotating magnetic fields, either driven by the disk or the black hole itself, can produce
energetic charged particles emerging from the vicinity of the black hole.
\subsection{Direct electrostatic field acceleration}
If a black hole is embedded in a poloidal field of strength $B_p$ and rotating with 
angular frequency $\Omega_H$, it will induce an electric field of magnitude $|\vec{E}| 
\sim (\Omega_H r_H) B_p/c$. This corresponds to a voltage drop across the horizon 
$r_H$ of magnitude $\Phi \sim r_H |\vec{E}|$. In terms of the electric circuit analogy,
a rotating black hole thus behaves like a unipolar inductor (battery) with non-zero 
resistance, so that power can be extracted by electric currents flowing between its 
equator and poles. Using parameters appropriate for Cen~A, the voltage drop is of 
the order of
\cite{tho86} 
\begin{equation}
  \Phi \sim 3 \times 10^{19} a \left( \frac{m_{\rm BH}}{10^8 M_{\odot}} \right) 
           \left(\frac{B_p}{10^4\mathrm{G}}\right)~~\mathrm{[V]}\,,
\end{equation} where $0\leq a\leq 1$ denotes the dimensionless Kerr parameter. If 
a charged particle (with charge number $Z$) can fully tap this potential, particle 
acceleration to ultra-high energies 
\begin{equation}
E=Z\,e\,\Phi \sim 3\times 10^{19} Z~\mathrm{eV} 
\end{equation} may become possible. This would suggest a rather heavy composition 
instead of a light one (e.g., iron nuclei instead of protons) for cosmic ray events above 
$E_c = 5 \times 10^{19}$ eV. Yet, whether such energies can, in fact, be achieved, 
seems questionable:  (i) In the plasma-rich environment of AGN (where the typical 
charge number density is much larger than the Goldreich-Julian one), a non-negligible 
part of the presumed electric field is expected to be screened and therefore not available 
for particle acceleration. 
(ii) Even if this would not be the case, curvature losses would constrain achievable 
proton energies in sources like Cen~A to values of $\lsim 10^{19}$ eV \cite{lev00}. 
(iii) Large-scale poloidal fields threading the horizon with strengths of $B_p \sim 
10^4$ G would be required. This may appear overly optimistic, at least in the case 
of a standard disk \cite{liv99}. (iv) A highly spinning black hole with $a \sim 1$ would 
be required (also, if one wishes to account for the power output solely via a 
Blandford-Znajek-type-process), although rather modest spins may be expected 
for FR~I sources \cite{dal09}. Taken together, this suggests that direct acceleration 
of protons to energies of $E_c$ and beyond in Cen~A is rather unlikely, while it could 
be (marginally) possible for heavy elements. 

\subsection{Centrifugal acceleration}
Even if the charge density would be such that effective electric field screening does
occur, particle acceleration due to inertial effects (i.e., centrifugal acceleration along
rotating magnetic fields) could still be possible, e.g. \cite{osm07,rie08}. The requirement 
that the associated acceleration timescale remains larger than the inverse of the 
relativistic gyro-frequency, however, then implies a maximum Lorentz factor for cosmic
rays \cite{rie08}, which in the case of Cen~A is of the order of 
\begin{equation}
  \gamma \lsim 2 \times 10^7 \gamma_0^{1/3} Z^{2/3} 
                  \left(\frac{m_p}{m_0}\right)^{2/3}
                  \left(\frac{r_{\rm L}}{10^{14}\mathrm{cm}}\right)^{2/3}\,
\end{equation} where $r_{\rm L}$ denotes the light cylinder radius (typically of the order 
of a few times the Schwarzschild radius). This suggests that centrifugal acceleration in 
Cen~A will be unable to account for the production of ultra-high energy cosmic rays.\\

\section{\large Fermi-type particle acceleration in the jets and beyond}
Stochastic processes in the turbulent AGN environment (e.g., in the jets or lobes) 
could well lead to the production of non-thermal particle distributions. In the classical 
Fermi picture \cite{fer49}, for example, particle acceleration occurs as a 
consequence of multiple scattering off moving magnetic turbulence structures, with 
a small energy change in each scattering event. The characteristic energy gain per 
scattering event for an energetic charged particle (velocity $v \sim c$), elastically 
scattering off some magnetic irregularity moving with typical velocity $\vec{u}$, is 
given by
\begin{equation}\label{gain}
\Delta \epsilon:=
\epsilon_2 -\epsilon_1 = 2 \Gamma^2 (\epsilon_1 u^2/c^2
                         -\vec{p}_1 \cdot \vec{u})\,,
\end{equation} where $\Gamma=(1-u^2/c^2)^{-1/2}$ is the Lorentz factor of the 
scatterer, $\vec{p}=\epsilon \vec{v}/c^2$ the particle momentum and the indices 1 and 
2 denote particle properties before and after scattering. A particle can thus gain or 
lose energy depending on whether it suffers head-on/approaching ($\vec{p_1} 
\cdot \vec{u} < 1$) or following/overtaking ($\vec{p_1} \cdot \vec{u} > 1$) collisions.
Based on this, one can distinguish the following Fermi-type particle acceleration 
processes, cf. \cite{kir01,duf06,rie07}. 

\subsection{Diffusive shock (Fermi I) acceleration}
Diffusive shock acceleration assumes that energetic particles (with gyro-radius 
much larger than width of the shock, $|\vec{p}_1|\simeq \epsilon_1/c$) can pass 
unaffected through a shock, and, by being elastically scattered in the fluid on either 
side, cross and re-cross the shock several times. Sampling the difference $\Delta 
u$ in flow velocities across a shock (always head-on), the characteristic energy 
gain for a particle crossing the shock, cf. eq.~(\ref{gain}), becomes first order in 
$\Delta u/c$, i.e., $\Delta \epsilon/\epsilon_1 \propto (\Delta u/c)$. As this energy gain 
is acquired during a shock crossing time $t_c \sim \lambda/u_s$ (with $u_s$ the 
shock speed and $\lambda$ the scattering mean free path), the characteristic 
acceleration timescale (for a non-relativistic shock) is of the order of 
\begin{equation}\label{tshock}
t_{\rm acc} \sim \frac{\epsilon}{(d\epsilon/dt)} \sim \frac{(\epsilon_1/\Delta\epsilon)}{t_c} 
                    \sim \lambda \frac{c}{u_s^2}\,.
\end{equation} If radiative losses are negligible, we can equate the timescale for acceleration 
with the one for cross-field diffusion out of the system, $t_e\sim r_w^2/(\lambda c)$, or 
the dynamical timescale, $t_d \sim z/u_s$ (whichever is smaller), to derive an estimate 
for the maximum achievable particle energy, cf. \cite{hil84}
\begin{equation}\label{hillas}
 E_{\rm max} \sim Z e B r _w \beta_s 
                        \sim 2\times 10^{19} Z \left(\frac{B_0}{10^4\mathrm{G}}\right)
                        \left(\frac{\beta_s}{0.1}\right)~~\mathrm{eV}\,,
\end{equation} taking $\lambda \sim r_{\rm gyro}$ to be of the order of the gyro-radius, 
$\beta_s=u_s/c$, and $B(z) \sim 4~B_0~(r_g/z \alpha_j)$ (allowing for some magnetic field 
compression), with $B_0$ the field strength close to the black hole and $\alpha_j$ the
jet opening angle. The observed (radio) jet speeds in Cen~A are only mildly relativistic 
with $u_j \sim 0.5$ c. If representative for the general flow, then typical internal shock 
speeds (of the order of the relative velocity between colliding shells) are expected to 
be rather moderate with $\beta_s \sim 0.1$ or less. Such low shock speeds are as well 
suggested by the nuclear SED of Cen~A, showing an electron synchrotron peak below 
$10^{20}$ Hz (already assuming the 2nd peak to be due to synchrotron and not inverse 
Compton, cf. \cite{len08}): synchrotron-limited electron shock acceleration would imply 
a (magnetic field-independent) peak at $\sim 3 \times 10^{19} (\beta_s/0.1)^2$ Hz and 
thereby support rather modest shock speeds. Equation~(\ref{hillas}) suggests that if shock 
acceleration would be responsible for UHECR production in Cen~A, then the expected 
composition should be rather heavy, i.e., efficient shock acceleration of protons to energies 
of $\sim E_c$ and beyond seems unlikely (see also below, \S \ref{jet_power}). This might 
be compared with a recent analysis of the PAO measurements suggesting that the cosmic
ray composition becomes heavier towards the highest measured energies \cite{hop09}.

\subsection{Stochastic Fermi II acceleration}
According to eq.~(\ref{gain}), particle acceleration due to scattering off randomly 
moving magnetic inhomogeneities is accompanied by an average energy gain
which is second order in $(u/c)^2$. Efficient acceleration thus obviously requires 
that the velocity of the scatterers is sufficiently large. As the energy gain is acquired 
over a scattering time $t_s \sim \lambda/c$, the associated acceleration timescale 
is of the order of  
\begin{equation}\label{t2ndFermi}
   t_{\rm acc} \sim \left(\frac{c}{v_A}\right)^2 \frac{\lambda}{c}\,,
\end{equation} assuming that the scattering is due to Alfv\'{e}n waves moving with a 
speed $u=v_A=B/\sqrt{4\pi\rho}$. If we again neglect radiative losses, achievable 
particle energies are limited by escape via cross-field diffusion, resulting in an upper 
limit of
\begin{equation}\label{Eshock}
  E_{\rm max} \sim 2 \times 10^{19} Z \left(\frac{R}{100~\mathrm{kpc}}\right)
                          \left(\frac{v_A}{0.1~c}\right)
                           \left(\frac{B}{10^{-6}\mathrm{G}}\right)~~\mathrm{eV}\,,
\end{equation} on scales of $R\sim 100$ kpc appropriate for the giant radio lobes in 
Cen~A. For relativistic Alfv\'{e}n speeds ($\gsim 0.3$ c), 2nd order Fermi effects could 
thus potentially allow proton acceleration up to ultra-high energies \cite{har09}. Yet, 
whether such conditions could be realized seems questionable, cf. \cite{osu09}. 
For if some of the observed X-ray emission in the giant lobes of Cen~A is indeed 
thermal in origin, e.g., \cite{iso01}, this would imply a thermal plasma density of the 
order of $n_{\rm th} \sim (10^{-5} - 10^{-4})$ cm$^{-3}$, so that expected Alfv\'{e}n 
speeds would be of the order of $v_A \sim 0.003$ c, i.e., well below the ones required.
Such (relatively high) thermal plasma densities are in fact consistent with recent, 
independent estimates based on Faraday rotation measurements in the radio lobes of 
Cen~A \cite{fea09}. Given current evidence, it may thus seem rather doubtful whether 
efficient UHECR acceleration could take place in its giant radio lobes. 

\subsection{Shear acceleration}
If the flow, in which the scatterers are thought to be embedded, has a smoothly 
changing velocity profile in the direction perpendicular to the jet axis (e.g., a shear 
flow or layer with $\vec{u}=u_z(r)\vec{e}_z$), then energetic particles, scattered 
across it, may well be able to sample the flow difference $du$ and thereby get 
accelerated \cite{jok90,rie06}. Like stochastic 2nd 
order Fermi acceleration, the average energy gain would be proportional to 
$(du/c)^2$, although the physical origin is now different (i.e., due to the systematic, 
instead of the random motion of the scatterers). The velocity difference in the flow, 
experienced by a particle scattered across it, is of the order of $du \sim (du_z/dr)
\lambda$, where $\lambda$ is the scattering mean free path. Again, this energy 
change is acquired over a mean scattering time $\tau_s \sim \lambda/c$, so that 
the characteristic acceleration timescale becomes
\begin{equation}
  t_{\rm acc} \sim \frac{\epsilon_1}{\Delta\epsilon/\tau_s} 
                      \sim \frac{1}{(du_z(r)/dr)^2}\frac{c}{\lambda}\,.
\end{equation} Compared to eq.~(\ref{tshock}) and eq.~(\ref{t2ndFermi}), the acceleration 
timescale is now inversely proportional to $\lambda$. Thus, as a particle increases 
its energy (so that the mean free path $\lambda$ becomes larger), the acceleration 
timescale decreases. Shear acceleration will, therefore, preferentially pick up high 
energy seed particles for further energization, and act more easily on protons than
on heavier nuclei. It seems well possible that shocks, operating in the jet (either on 
smaller scales or within a spine), could provide the energetic seed protons required 
for further shear acceleration along the jet \cite{rie09}. If so, then the maximum 
achievable energies might be expected to be essentially determined by the 
confinement condition that the gyro-radius remains smaller than the width of the 
shear layer. The large-scale jet in Cen~A has a projected length of $\sim 4.5$ kpc 
and towards its end a width of about $\sim 1$ kpc \cite{bur83,kra02}. If we take a 
characteristic magnetic field strength of $B\sim 10^{-4} b_j$ G on kpc-scale and 
assume the width of the shear to become comparable to the width of the jet, 
achievable maximum energies would be of the order of
\begin{equation}
E \sim Z e B (\Delta r) \sim 10^{20} b_j Z ~\mathrm{eV}\,,
\end{equation} suggesting that shear acceleration might be able to boost energetic 
seed protons (produced by shock acceleration) up to energies beyond $E_c$. Note 
that in the presence of sufficient internal shear, the magnetic field within the layer 
may well be expected to fall more slowly with distance along the jet, $b_j \gsim 1$, 
due to amplification by stretching and folding of magnetic field lines, e.g. 
\cite{beg84,urp06}. A shear dynamo effect could possibly also explain why in
Cen~A the magnetic field direction seems to be almost parallel along the kpc jet  
\cite{har03}. If such an amplification takes place, the situation may be even more 
favorable.\\ 

\section{\large Constraints from jet power requirements}\label{jet_power}
If efficient UHECR acceleration would take place in the jet of Cen~A, one could 
estimate the magnetic energy flux carried by the jet, and therefore the minimum 
jet power required. For the magnetic flux carried by the jet in Cen~A, we have 
\begin{equation}
 L_m \sim  \int dr~2 \pi r u_z (B_{\perp}^2/8\pi)  = r^2 B_{\perp}^2 u_z/8 \,,
\end{equation} where $B_{\perp}$ is the magnetic field component perpendicular 
to the direction of the bulk outflow velocity $u_z$ (assumed to be non-relativistic), 
and where the second equality holds provided $B_{\perp}$ and $u_z$ (or more 
precisely, the product $B_{\perp}^2 u_z$) are independent of the jet radius $r$. 
If we assume $r \sim r_w/2$ and use eq.~(\ref{Eshock}) to find an expression for 
the magnetic field in terms of $E_{\rm max}$, efficient cosmic ray acceleration by 
internal shocks would require a jet power of at least $L_j \sim 2 L_m$, i.e.
\begin{equation}
L_j \sim 10^{44} \left(\frac{u_z}{0.5 c}\right) 
                                            \left(\frac{0.1}{\beta_s}\right)^2   
                                           \left(\frac{E_{\rm max}}{10^{19}\mathrm{eV}}\right)^2
                                           \frac{1}{Z^2}~~\mathrm{erg/s}\,. 
\end{equation} This would support the previous conclusion that proton acceleration 
beyond a few times $10^{19}$ eV would require a jet power well in excess of the one 
expected for Cen~A as an FR~I source, e.g., \cite{ghi01}. On the other hand, UHECR 
acceleration of heavy elements like iron may still remain possible. In the case of 
shear acceleration, the parameters employed above for efficient proton acceleration 
($B\sim10^{-4}$ G, $r\sim0.5$ kpc, $u_z \sim0.5$ c) may, at first sight, as well imply 
a jet power of $L_j \sim r^2 B_{\perp}^2 u_z/4 \sim 10^{44} (E_{\rm max}/10^{20}
\mathrm{eV})^2 Z^{-2}$ erg/s. However, this ignores the $r$-dependence of the bulk 
flow and (probably) the magnetic field, and when properly accounted for, a smaller 
jet power may already well be sufficient.\\

\section{\large Conclusions}
The above analysis suggests that efficient acceleration of protons to UHECR energies
in Cen~A  is challenging and may require the operation of an additional acceleration
mechanism like shear to further boost achievable particle energies beyond $E_c = 5 
\times 10^{19}$ eV. Efficient shear acceleration in Cen~A would require high energy 
seed particles which, however, could be provided by, e.g., shock acceleration. A 
fraction of these seed protons may then be picked up and accelerated to the maximum 
energy given by the confinement limit. If such a two-step process would indeed take 
place, spectral changes in the cosmic ray energy spectrum may not just simply be due 
to propagation effects. The situation is much more relaxed for heavier elements like iron 
nuclei, which could possibly be directly accelerated (either by shocks or within the black 
hole magnetosphere) to energies of $E_c$ and beyond. If Cen~A would indeed be an 
efficient UHECR accelerator one may thus expect the composition to become
heavier above energies $\sim 10^{19}$ eV.\\

\section*{\large Acknowledgement}
Most of this work was presented at the 4th International JEM-EUSO Workshop (Torino, 
December 2008). The author is very thankful to the organizers for invitation, and to the 
participants for inspiring discussions. Discussions with Martin Hardcastle, Ricard 
Tomas and Sergey Troitsky during the Trondheim SOCoR Workshop 2009, and 
helpful comments on the manuscript by Andrew Taylor are gratefully acknowledged.\\



\end{document}